\documentstyle[aps,preprint,epsf]{revtex}
\newcommand{\GeV}{~\mbox{GeV}}
\newcommand{\Gev}{~\mbox{GeV}}

\begin{document}
\draft

\preprint{CERN-TH/2000-263}
\title{Intercepts of the non-singlet
structure functions}

\author{B.I. Ermolaev\\
A.F. Ioffe Physico-Technical Institute, 194021 St.Petersburg, Russia\\
M. Greco \\
CERN, 1211 Geneva 23, Switzerland \\
 and\\
Dipartimento di Fisica and INFN, University of Rome III, Rome, Italy \\
S.I. Troyan\\
Petersburg Nuclear Physics Institute, 188300 Gatchina, St.Petersburg,
Russia}

\maketitle

\begin{abstract}
Infrared evolution equations for small-$x$ behaviour of
the non-singlet structure functions $f_1^{NS}$ and $g_1^{NS}$
are obtained
and solved in the next-to-leading approximation, to all orders
in $\alpha_s$, and including
running $\alpha_s$ effects. The intercepts  of these structure
functions, i.e. the
exponents of the  power-like small-$x$ behaviour, are calculated.
A detailed comparison with  the leading logarithmic approximation (LLA)
and DGLAP is made.
We explain why the LLA predictions for the small-$x$ dependence
of the structure functions may be more reliable than the prediction for
the $Q^2$ dependence in the range of $Q^2$ explored at HERA.
\end{abstract}




\section{Introduction}

In recent years the deep inelastic
scattering (DIS) of leptons off hadrons has been the object of
intensive theoretical studies.
In QCD the hadronic tensor for electron proton DIS is
usually regarded as a
convolution of two objects: the ``partonic'' tensor $W_{\mu\nu}$
describing DIS
off a (nearly) on-shell parton (quark or gluon) and the
probability $P_h^p$ to find the parton in
the nucleon. As is well known, they have a different description: whereas
the tensor $W_{\mu\nu}$ can be calculated within
perturbation theory, there are no model-independent methods
of calculating $P_h^p$, but only the general behaviour
with respect to the kinematical variables can be predicted.
Usually, the hadronic tensor $W_{\mu\nu}$ is
presented in
terms of the structure functions (see e.g.  \cite{f}) $f_{1,2}$ and
$g_{1,2}$:

\begin{eqnarray}\label{totwmunu}
W_{\mu\nu} &=&
\left(-g_{\mu\nu}+\frac{q_{\mu}q_{\nu}}{q^2} \right) f_1 +
\left( p_{\mu} - \frac{(pq)}{q^2}q_{\mu} \right)
\left( p_{\nu} - \frac{(pq)}{q^2}q_{\nu} \right) \frac{1}{(pq)} f_2
\nonumber\\ &+& \imath\epsilon_{\mu\nu\lambda\sigma}\frac{m
q^{\lambda}}{(pq)} \left[ S^{\sigma} g_1 +
\left( S^{\sigma}-\frac{(Sq)}{pq} p^{\sigma} \right) g_2 \right],
\end{eqnarray}
where $p$, $m$, $S$ are the momentum, mass and polarization of the
incoming parton and $q$ is the momentum of the virtual photon.
The structure functions depend on the Bjorken variable
$x=-q^2/(2pq)$, with $(1>x>0)$ and  $Q^2=-q^2>0$.
Each of the structure functions consists of flavour singlet and
flavour non-singlet (NS) contributions. Although the singlet
contributions accounting for gluon splitting are dominant at small $x$,
the NS contributions are also interesting quantities to investigate.
They can be expressed in terms of densities of quarks ($\Delta
q$) and antiquarks ($\Delta\bar{q}$) inside a
hadron.
The non-singlet spin-dependent structure function $g_1^{NS}$
refers to the combination of
($\Delta q+\Delta\bar{q}$), whereas the non-singlet
spin-averaged structure
function $f_1^{NS}$ is related to  ($\Delta q-\Delta\bar{q}$);
combining them one could study the evolution of $\Delta q$ and
$\Delta\bar{q}$ separately.

The standard theoretical investigation
of DIS structure functions, originally developed for the kinematic
region of $x\sim1$ and large $Q^2$,  is made through the DGLAP
\cite{dglap}  evolution
equations
. This approach accounts for the evolution of the
structure
functions with respect to the photon virtuality $Q^2$, whereas the
evolution
with respect to $x$ is neglected. In other words, the DGLAP accounts
for the resummation of all powers of $\ln Q^2$ and systematically neglects
the powers of $\ln x$,
which is obviously correct  when $x$ is not
small.  Despite the fact that the DGLAP equations provide a good agreement
with
experimental data (see e.g. \cite{a}), they are not expected to work
well in the small-$x$ region, where logarithms of $x$ are not less
important than logarithms of $Q^2$ and therefore must be taken into
account to all orders in $\alpha_s$.  This resummation leads to a
power-like, or Regge-like, behaviour in $x$ and in $Q^2$ of the type
$f_1^{NS}\propto(\sqrt{Q^2}/x)^a$ .  Such a behaviour cannot
be obtained within DGLAP, which predicts a dependence of the type
$f_1^{NS}\propto\exp\sqrt{c\ln(1/x)\ln\ln Q^2}$.

In perturbative QCD the power-like behaviour was obtained
long ago  \cite{bfkl} for the singlet structure functions of
spin-averaged
DIS and later in refs. \cite{emr,ber} for the structure functions $g_1$
and $f_1^{NS}$.  However, contrary to the DGLAP, the results of
refs. \cite{bfkl,emr,ber} are obtained in the leading logarithmic
approximation (LLA) with a fixed QCD coupling
$\alpha_s$. Hence the exponents calculated in
\cite{bfkl,emr,ber} (also called intercepts) for the predicted
power-like small-$x$ behaviour of all structure functions contain
$\alpha_s$ fixed at a scale that is not well defined. The very existence
of such a scale, sometimes called ``a reasonable scale'', and
estimates for its numerical value have been discussed in many
works (see e.g. \cite{bfklp}).
However, according to the
Regge theory, the expressions for the intercepts must not contain
$\alpha_s$ running with $Q^2$, but, on  the contrary,
they just have to be numbers.

Thus, in order to obtain realistic values for the intercepts, one has to
go
beyond the LLA and include the QCD running coupling effects in the
evolution equations
at small $x$, so that the running coupling $\alpha_s$ would be integrated
out in the solutions.
Recently in ref. \cite{egt} we have obtained expressions for the
non-singlet
structure  function\footnote{$f_1^{NS}$ was denoted as $f^{NS}$ in work
\cite{egt}.} $f_1^{NS}$, with running QCD coupling effects taken into
account.  The resulting intercept does not explicitly depend on
$\alpha_s$ but depends on the infrared cut-off.
That calculation prompted us to study the running QCD coupling effects
for the non-singlet structure functions in more detail.

The structure function $g_1^{NS}$ is similar to $f_1^{NS}$, because
the same Feynman graphs contribute to both of them. As a result,
they obey the same leading order (LO) DGLAP evolution
equation. A difference between them arises only in next-to-leading
order (NLO) corrections to DGLAP and can be extracted from
the expressions of the second loop anomalous dimensions
(see e.g. \cite{grsv,fkl}).

Concerning the small-$x$ behaviour, the difference between $f_1^{NS}$
and $g_1^{NS}$ can be related to the difference in their
signatures: the signature of $f_1^{NS}$ is positive whereas $g_1^{NS}$
has a negative one. A  difference between them means
that some high-order non-ladder graphs contribute to $f_1^{NS}$ and
$g_1^{NS}$ in a different way.  Historically, signatures for scattering
amplitudes of high energy processes were first introduced in the
context of the Regge theory (see e.g. \cite{col}). A
forward scattering amplitude $M(s,u)$, where $s$ and $u$ are
the Mandelstam variables, has positive (negative) signature if it
is symmetrical (antisymmetrical) with respect to the replacement
$s\leftrightarrow u$:

\begin{equation}
\label{sign}
M^{(+)}(u,s) = M^{(+)}(s,u),~~~~M^{(-)}(u,s) = -M^ {(-)}(s,u) .
\end{equation}
Basically, any forward
scattering amplitude $M(s)$ consists of the two parts:

\begin{equation}
\label{ms}
M(s,u) = M^{(+)} +
M^{(-)},~~~~M^{(\pm)}=\frac12\left(M(s,u)\pm M(u,s)\right) .
\end{equation}
In  the small-$x$ region, neglecting $O(m^2/s)$ and $O(Q^2/s)$ terms, one
can use $u\approx-s$. As the imaginary part in $s$, $\Im_s M$, for
$s > 0$, corresponds to the cross-section, this  cross-section may
acquire contributions from both $M^{(+)}$ and $M^{(-)}$ amplitudes.
The partonic tensor $W_{\mu\nu}$ of Eq. (\ref{totwmunu}) can be considered
as the imaginary part of the amplitude $M_{\mu\nu}$ of the forward Compton
scattering
of a virtual photon on a constituent quark:
\begin{eqnarray}\label{mmunu}
M_{\mu\nu} &=&
\left(-g_{\mu\nu}+\frac{q_{\mu}q_{\nu}}{q^2} \right) 2M_1 +
\left( p_{\mu} - \frac{(pq)}{q^2}q_{\mu} \right)
\left( p_{\nu} - \frac{(pq)}{q^2}q_{\nu} \right) \frac{1}{(pq)} 2M_2
\nonumber\\ &+& \imath\epsilon_{\mu\nu\lambda\sigma}\frac{m
q^{\lambda}}{(pq)} \left[ S^{\sigma} 2M_3 +
\left( S^{\sigma}-\frac{(Sq)}{pq} p^{\sigma} \right) 2M_4 \right],
\end{eqnarray}
so that

\begin{equation}
\label{fm}
W_{\mu\nu}=\frac{1}{2\pi}\Im_s M_{\mu\nu},
\end{equation}

and in particular,
\begin{equation}
\label{fmi}
f_1 =\frac{1}{\pi}\Im_s M_1,~~~~~~g_1 = \frac{1}{\pi}\Im_s M_3~.
\end{equation}

The amplitude
$M_{\mu\nu}$ is
invariant with respect to the replacement $q\to -q$,
$\mu\leftrightarrow\nu$\, which corresponds to the crossing transition
from the
$s$-channel to the  $u$-channel. However, the
projection operators multiplying $f_1$ and $g_1$ behave differently
under such a transition:  the operator
$(g_{\mu\nu} - q_{\mu}q_{\nu}/q^2)$ does not change, so the invariant
amplitude $M_1$ multiplying it does not either.
On the contrary, the operator  $(\imath\epsilon_{\mu\nu\lambda\sigma} m
q_\lambda/(pq))$ acquires a negative sign, so the invariant amplitude
$M_3$ must acquire a  negative sign too.  Therefore,
the invariant amplitudes

\begin{equation} \label{mpos}
M^{(+)} = \frac12\left(M_1(s,u) + M_1(u,s)\right)
\end{equation}
and

\begin{equation}
\label{mneg}
 M^{(-)} \equiv\frac12\left(M_3(s,u) - M_3(u,s)\right)
\end{equation}
have positive and
negative signatures,
respectively, and the DIS structure
functions $f_1$ and $g_1$ can be considered as imaginary parts of
scattering amplitudes with positive and negative signatures:

\begin{equation}\label{ims}
f_1 = \frac{1}{\pi}\Im_s M^{(+)}~,\qquad g_1 = \frac{1}{\pi}\Im_s
M^{(-)}~.
\end{equation}

The first observation of the
importance of the contribution of the negative-signature amplitudes to
the cross-sections was made in the context of QED in ref. \cite{gln}.
Later on, in ref. \cite{kl}, quark-quark scattering amplitudes with
positive and negative signatures were calculated in the
double-logarithmic approximation (DLA). Using the results of those works,
it was shown in ref. \cite{ber} that in DLA, with $\alpha_s$ fixed, the
intercept of $g_1^{NS}$ is larger  than the intercept of $f_1^{NS}$.

In the present paper we
obtain explicit expressions for $f_1^{NS}$ and $g_1^{NS}$ for
small values of $x$,
which account for both
leading double-logarithmic (DL) and sub-leading single-logarithmic (SL)
contributions to all orders in QCD coupling, including
running $\alpha_s$ effects.
As logarithms of both $x$ and $Q^2$ are
important at small $x$, we account for both of them, constructing
and solving a two-dimensional infrared evolution equation (IREE).
In order to take  running $\alpha_s$ effects into account, we use the
approach of ref. \cite{kl}, with the improvement made in our previous work
\cite{egt}. Also, we discuss similarities and differences between
running $\alpha_s$ effects in the DGLAP and in our
approach. Finally we discuss the region of applicability of the LLA
and estimate the range of $Q^2$ where it can be safely applied.
The paper is organized as follows:
in Section~2 we derive the small-$x$ evolution equations  for
the non-singlet structure functions
and solve them.
In Section~3 we obtain the
asymptotic behaviour of $f_1^{NS}$ and $g_1^{NS}$.
In Section~4 we compare our results with the
intercepts of the non-singlet
structure functions obtained in  LLA.
In Section~5 our findings are compared with DGLAP
predictions for the NS anomalous dimensions and we discuss the
difference in accounting for running $\alpha_s$ effects.
Finally, Section  6
contains our conclusions.

\section{Small-$x$ evolution equation for the non-singlet
structure functions}

\begin{figure}
\begin{center}
\begin{picture}(300,150)
\put(0,10){
\epsfbox{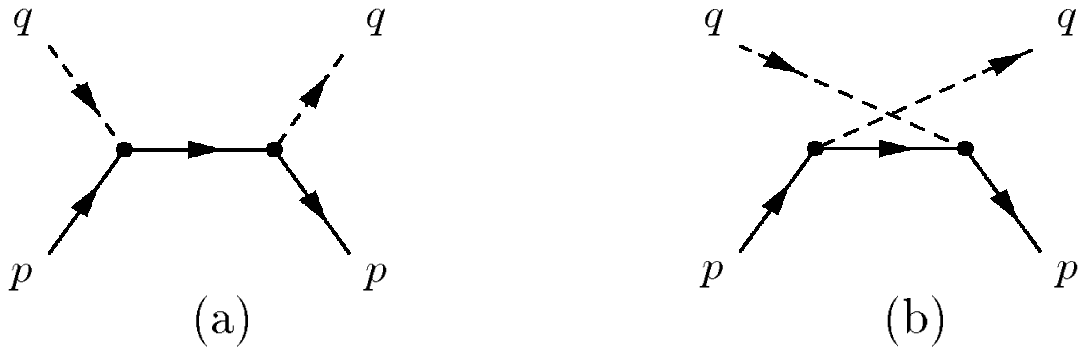}
}
\end{picture}
\end{center}
\caption{The Born graphs for the DIS amplitude $M_{\mu\nu}$.}
\label{Born}
\end{figure}

In the Born approximation, which is the pure NS case, $M_{\mu\nu}$
is given by the sum of the
graphs in Fig.~\ref{Born}. They yield

\begin{equation} \label{mmunuborn}
M^{Born}_{\mu\nu} = \left(-g_{\mu\nu} +
2x\frac{p_{\mu}p_{\nu}}{(pq)}\right) 2M^{(+)}_{Born} + \left(
\imath\epsilon_{\mu\nu\lambda\sigma} \frac{mq^{\lambda}}{(pq)}
S^{\sigma} \right) 2M^{(-)}_{Born},
\end{equation}
where
\begin{equation} \label{mborn}
M^{(\pm)}_{Born} = e^2_q \frac{1}{2}\left[ \frac{s-m^2-q^2}{m^2
-s-\imath\epsilon} \pm
\frac{u-m^2-q^2}{m^2-u-\imath\epsilon} \right],
\end{equation}
with  $e_q$ being the
electric charge of the incoming quark
and $s=(p+q)^2$, $u=(p-q)^2$  the Mandelstam variables.
We  have omitted unimportant contributions proportional
to $q_{\mu}$, $q_{\nu}$ in Eq.~(\ref{mmunuborn}).

$M^{(\pm)}_{Born}$ are amplitudes of defined
signature (cf. Eq.~(\ref{sign})).
Obviously, only the first term in Eq.~(\ref{mborn}), corresponding to
the graph (a) in Fig.~\ref{Born}, has an imaginary part in $s$ at $s >
0$, which gives the same contribution to the amplitudes of both
signatures.  According to Eq.~(\ref{ims}) we obtain the well-known
results:

\begin{equation} \label{fgborn}
f_1^{Born} = g_1^{Born} = \frac{e^2_q}{2}\delta(1-x)~.
\end{equation}

\begin{figure}
\begin{center}
\begin{picture}(300,180)
\put(0,0){
\epsfbox{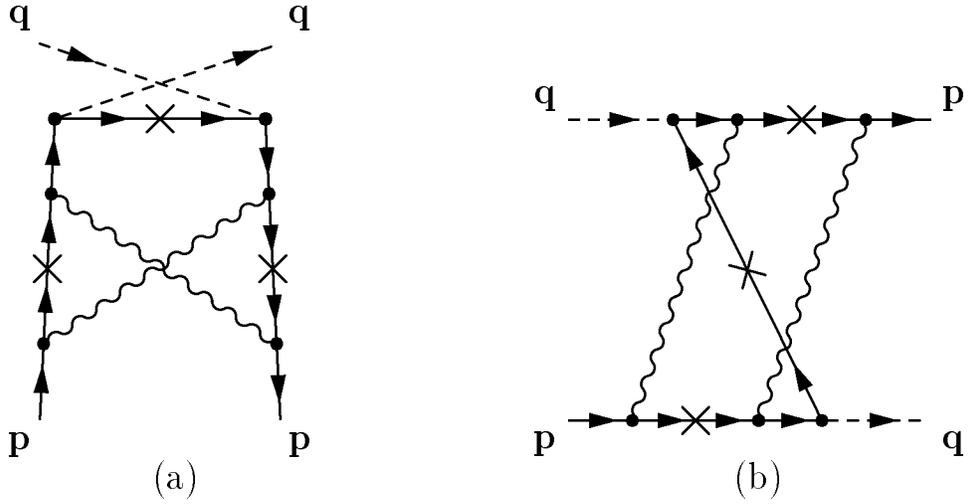}
}
\end{picture}
\end{center}
\caption{Two-loop graph obtained from Fig.~\ref{Born}b, contributing
to $\Im_s M_{\mu\nu}^{(-)}$ (a), and the corresponding physical
process (b).}
\label{second}
\end{figure}

The property $f_1^{Born} = g_1^{Born}$ remains true in the next one-loop
 approximation, with
one gluon added to graphs in Fig.~\ref{Born}. Only the dressing of
graph (a) in Fig.~\ref{Born} contributes to the imaginary part in $s$
at $s>0$. This explains why the LO DGLAP splitting functions and the
LO anomalous dimensions are the same for the structure functions
corresponding to amplitudes with different signatures.  The difference
between them arises only at the second-loop order. Indeed, let us
consider graph (a) of Fig.~\ref{second}. It is actually a
non-ladder graph, obtained from the Born graph (b) in Fig.~\ref{Born} by
adding two gluons to it. On the other hand, graph (a) in
Fig.~\ref{second} can be redepicted as graph (b) in Fig.~\ref{second}.
It is easy to check that its imaginary  part in $s$ (the quark
propagators to be cut are marked with crosses in Fig.~\ref{second})
has, in particular, DL terms ($\sim \ln^3(1/x)$);
this graph therefore cannot be neglected.  Then, one can see that the DL
contribution of this graph, which is symmetrical in $\mu,\nu$
(contribution
to $f_1^{NS}$) and the antisymmetrical one (contribution to $g_1^{NS}$)
have different signs. This example shows that some higher-loop
contributions to $f_1^{NS}$ and $g_1^{NS}$ are different.

\begin{figure}
\begin{center}
\begin{picture}(300,150)
\put(0,10){
\epsfbox{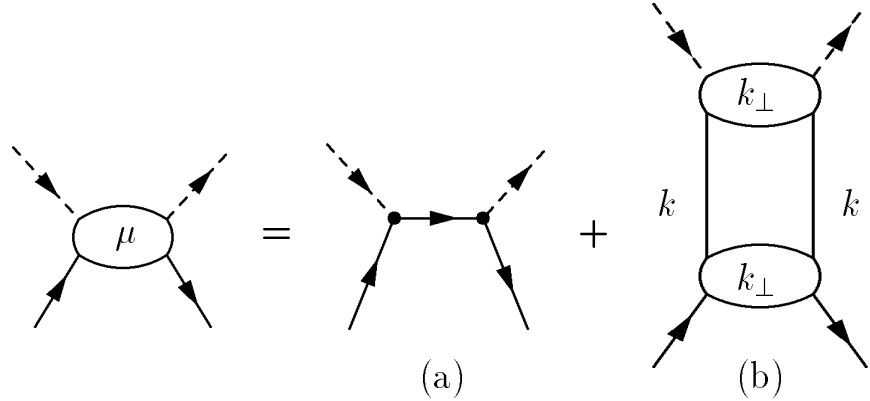}
}
\end{picture}
\end{center}
\caption{The evolution equation for the DIS amplitude $M_{\mu\nu}$.}
\label{equation}
\end{figure}

Thus, a possible regular way  of
calculating $f_1^{NS}$ and $g_1^{NS}$ consists  of the following steps:
\begin{itemize}
\item[(i)] Calculate the forward scattering Compton
amplitude $M_{\mu\nu}$ obtained by adding gluon propagators
to both Born graphs in
Fig.~\ref{Born}.
\item[(ii)] Extract  from it the amplitudes $M_1$ and $M_3$
proportional to
$(-g_{\mu\nu})$ and
$\imath\epsilon_{\mu\nu\lambda\sigma}q^{\lambda}p^{\sigma}/(pq)$ (cf.
Eqs.~(\ref{mmunu}) and (\ref{fmi})), which are the positive- and negative-
signature amplitudes $M^{(\pm)}$, respectively.
\item[(iii)] Calculate $(1/\pi)\Im_s M^{(\pm)}$ .
\end{itemize}
So, first  we calculate $M_{\mu\nu}$, and we do it by constructing
and solving an IREE for $M_{\mu\nu}$.
In order to account for the evolution in both $x$ and $Q^2$
one must consider DL and SL contributions to all orders in
$\alpha_s$.  High order Feynman graphs contributing to $M_{\mu\nu}$
may have
both ultraviolet (UV) and infrared (IR) singularities. Whereas UV
singularities are absorbed by renormalization, IR ones must be
regulated explicitly.  Providing gluons with a mass violates the
gauge invariance. In order to save it and to avoid IR singularities, we
use the Lipatov prescription of compactifying the impact parameter space
(see e.g. works \cite{l,kl,el}). In other words, we introduce the
infrared cut-off $\mu$ in the transverse space (with respect to the
plane formed by $q$ and $p$) for integrating over the momenta of
virtual particles:

\begin{equation}
\label{mu}
 k_{i\perp} > \mu~.
\end{equation}
With such a cut-off acting as a mass scale, one can neglect quark
masses and still be free from IR singularities.  As a result the
amplitude $M_{\mu\nu}$ depends on the cut-off and we can study its evolution
with respect to it, thus obtaining the
IREE for $M_{\mu\nu}$.  Feynman graphs contributing to the
non-singlet structure functions are obtained from graphs in Fig.~\ref{Born}
by adding gluon propagators to them, without breaking the
quark lines, however.  With logarithmic accuracy, the  region of
integration
over transverse momenta of virtual quarks and gluons can be regarded
for every Feynman graph contributing to $M_{\mu\nu}$ as a sum of
sub-regions so
that in every sub-region only one virtual particle (quark or
gluon) has the minimal transverse momentum $k_{\perp}$. We call
such particles the softest ones although their longitudinal momenta can be
large (see e.g. \cite{el}). It is essential that integrating
over $k_{\perp}$ one
includes $\mu$ as the lowest limit. The softest particle can be
either a gluon or a
quark.
When the softest
particle is a quark, DL contributions come from the
sub-region where another quark has the same transverse momentum, so that
there appears a $t$-channel intermediate state with a soft
quark pair. Therefore
in order to keep the DL contributions, one must consider
$M_{\mu\nu}$  as convolution of two amplitudes
connected by the softest quark pair, each quark with the minimal
transverse momentum
$k_{\perp}$, as depicted by graph (b) in
Fig.~\ref{equation}.
The first amplitude appearing in graph (b) (the upper blob) is the Compton
forward
scattering
$M_{\mu\nu}$  and the second
one (the lower blob) is the forward scattering amplitude of quarks
$M_0$. On the other hand,
when the particle with the minimal $k_{\perp}$ is a virtual gluon, it can
be factorized out of the amplitude with the DL accuracy \cite{efl}. This
means, in particular, that
the propagator of such softest gluon is attached to external lines only,
yielding zero result.
Having added the
Born contribution to the convolution graph (b) in Fig,~3, we arrive at the
equation of the Bethe - Salpeter type
for $M_{\mu\nu}$.
As usual the Bethe - Salpeter equation relates $M_{\mu\nu}$ with on-shell
quarks (lhs of equation in Fig.~\ref{equation}) to $M_{\mu\nu}$
with off-shell quarks (rhs of Fig.~\ref{equation}).
However,  all amplitudes in Fig.~\ref{equation} are actually on-shell
because the
virtual quark
pair in Fig.~\ref{equation} has the minimal transverse momentum $k_{\perp}$.
Integrations in blobs imply $k_{\perp}$ to be a new IR cut-off and,
therefore,
a new mass scale for these blobs (see \cite{kl,el,efl}).

Let us note that in the DGLAP approach such softest quark pair is always
in the lowest ladder rung because of
the DGLAP ordering: in this approach all ladder transverse momenta
$k_{i \perp}$ are indeed ordered as

\begin{equation}
\mu\leq k_{1 \perp}\leq k_{2 \perp}\leq ...\leq Q^2,
\label{order}
\end{equation}
the numeration running from the bottom of the ladder to the top.
Equation ~(\ref{order})
shows that only the integration over $k_{1\perp}$ has $\mu$
as the lower limit.
The lower limits for other $k_{i\perp}$ with $i \neq 1$
are expressed through the other transverse momenta. Therefore the
small-$x$
DGLAP equations correspond to the case where the quark scattering
amplitude in graph (b) in Fig.~\ref{equation} is considered in the Born approximation.
As is well known, the ordering of Eq.~(\ref{order}) corresponds to
accounting for $\ln^n(Q^2/\mu^2)$  without considering terms $\ln^m(1/x)$
not accompanied by logarithms of $Q^2$.  On the other hand, as we
investigate the small-$x$ region, we have to lift this ordering,
allowing for the transverse momentum of quarks in any ladder rung to
reach the lowest limit $k_{\perp} = \mu$.  Obviously, this increases
the region of integration over $k_{i \perp}$.
Then, a quark pair with minimal
$k_{\perp}$ appears in any rung of the ladder Feynman graph.

The convolution of the amplitudes depicted in graph
(b) prompts us to apply the Mellin transform for solving this equation.
However, to respect the signatures of $M^{(\pm)}$ it is more convenient
to use the asymptotic form of the Sommerfeld - Watson (SW) transform (see
\cite{col}):

\begin{equation}
M^{(\pm)}(\frac{s}{\mu^2}, \frac{Q^2}{\mu^2}) =
\int_{-\imath\infty}^{\imath\infty} \frac{d\omega}{2\pi\imath}
\left(\frac{s}{\mu^2}\right)^{\omega}\xi^{(\pm)}(\omega)F^{(\pm)}
(\omega, \frac{Q^2}{\mu^2}),
\label{mellin}
\end{equation}
where

\begin{equation}
\label{xi}
\xi^{(\pm)} = -\frac{e^{-\imath\pi\omega} \pm 1}{2}
\end{equation}
is the signature factor, for which this
transform differs from that of Mellin. The inverse transform to
Eq.~(\ref{mellin}) is

\begin{equation}
F^{(\pm)}(\omega) = \frac{2}{\sin(\pi \omega)}
\int_0^{\infty}d\rho \exp(-\omega \rho) \Im_s M^{(\pm)}(\rho),
\label{invmellin}
\end{equation}
where $\rho = \ln(s/\mu^2)$.

The small-$x$ region corresponds effectively to the dominance of the
small-$\omega$ region in Eq.~(\ref{mellin}). Expanding $\xi$ into
series in $\omega$ and retaining the first terms in Eq.~(\ref{mellin},
$\xi^{(+)}\approx -1+\imath\pi\omega/2$ and
$\xi^{(-)}\approx\imath\pi\omega/2$, we see that in the
small-$x$ (or small-$\omega$) region

\begin{equation}\label{defns}
\frac{1}{\pi}\Im_s M^{(\pm)} = \frac12
\int_{-\imath\infty}^{\imath\infty} \frac{d\omega}{2\pi\imath}
\left(\frac{s}{\mu^2}\right)^{\omega} \omega
F^{(\pm)}(\omega,\frac{Q^2}{\mu^2});
\end{equation}
we need to calculate $F^{(\pm)}$ only to obtain the structure
functions $f_1^{NS}$ and $g_1^{NS}$ (cf. Eq.(\ref{fmi})). To do this
one can apply the SW transform of Eq.~(\ref{mellin}) to the
equation depicted in Fig.~\ref{equation} and then differentiate it with
respect to $\ln\mu^2$. On the other hand the Born amplitude
does not depend on $\mu$ and therefore vanishes when differentiated
with respect to $\mu$. Then, observing that

\begin{equation}
-\mu^2 \frac{\partial M^{(\pm)}}{\partial\mu^2} =
\frac{\partial M^{(\pm)}}{\partial \ln(s/\mu^2)} +
\frac{\partial M^{(\pm)}}{\partial \ln(Q^2/\mu^2)}
\label{lhs}
\end{equation}
corresponds to

\begin{equation}
\omega F^{(\pm)} + \frac{\partial F^{(\pm)}}{\partial y}
\label{lhsmel}
\end{equation}
for the amplitudes $F^{(\pm)}$, where we have defined $y =
\ln(Q^2/\mu^2)$, and differentiating the rhs of the equation depicted in
Fig.~\ref{equation} with respect to $\ln\mu^2$, we are led to the
following
IREE :

\begin{equation} \label{eqir}
\left(\frac{\partial}{\partial y} + \omega \right) F^{(\pm)}(\omega, y)
= \left[\frac{1}{8\pi^2}(1 + \lambda \omega)\right] F^{(\pm)}(\omega,
y) F^{(\pm)}_0(\omega)~~,
\end{equation}
where $\lambda = 1/2$. For more details, see ref. \cite{egt}, where the
latter equation was obtained
for $F=F^{(+)}$.
The only difference now is that IREE (\ref{eqir})
involves also the negative signature amplitude $F_0^{(-)}$ of forward
quark - quark scattering instead of $F_0^{(+)}$ only. The solutions
to Eq.~(\ref{eqir}) are

\begin{equation}\label{cir}
F^{(\pm)}= C\exp\left( \left[ -\omega+(1+\lambda\omega)
F_0^{(\pm)}(\omega) /8\pi^2 \right] y \right)~~,
\end{equation}
which contain two unknown quantities. The first one
is an arbitrary factor $C$, essentially of a non-perturbative nature,
which can be fixed by comparison with the data.

\begin{figure}
\begin{center}
\begin{picture}(300,300)
\put(0,10){
\epsfbox{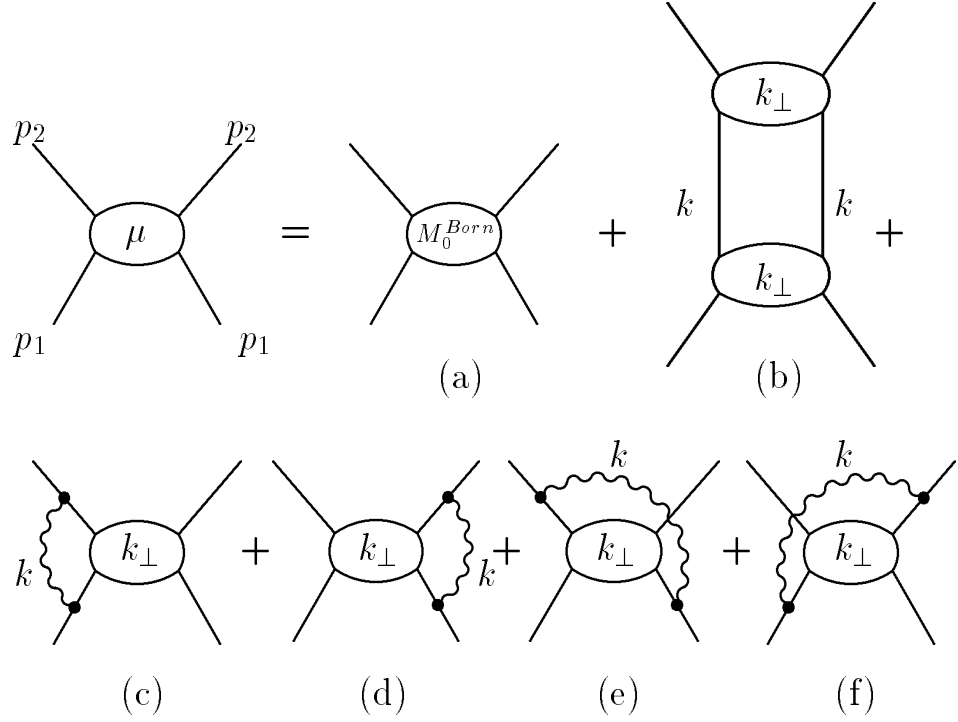}
}
\end{picture}
\end{center}
\caption{The evolution equation for the quark scattering amplitude
$M_0$.}
\label{f0}
\end{figure}

The other unknown quantity $F^{(\pm)}_0$ can be specified by
constructing and solving the IREE for the amplitude $M_0$ of forward
scattering of quarks.  The equation for $M_0$ is shown in
Fig.~\ref{f0}.  The first two terms in the rhs look similar to the
equation
in Fig.~\ref{equation}.  Indeed, the rhs consists of the Born
contribution depicted by graph (a), the convolution of two quark
scattering amplitudes (graph (b)), the intermediate quarks having
minimal transverse momenta $k_{\perp}$, and a new contribution depicted
by graphs (c) - (f).  For symmetry, the contributions of graphs (c)
and (d) are equal. The same is true for graphs (e) and (f).  All these
graphs correspond to the case where the particle with the minimal
transverse momentum $k_{\perp}$ is a gluon. The propagators of such
soft gluons are attached to external lines because by integrating over
the soft gluon momenta the most singular DL contributions come from the
region where each soft virtual gluon is factorized out (see e.g. ref.
\cite{efl}). Self-energy graphs are absent in Fig.~\ref{f0} because the
Feynman gauge is used. The blobs in these graphs imply integrations
over momenta of internal virtual particles with $k_{\perp}$ acting as a
new IR cut-off.
Let us consider in detail the contribution of the graphs in the rhs of
the equation
in Fig. \ref{f0}. The Born contribution
coming from graph (a) in Fig.~\ref{f0} is given by

\begin{equation}\label{defB}
B^{(\pm)} = -2\pi C_F\left[
\alpha_s(s) \frac{s}{s-\mu^2+\imath\epsilon} \pm
\alpha_s(-s) \frac{-s}{-s-\mu^2+\imath\epsilon}
\right],
\end{equation}
where the quark mass is substituted by $\mu$ in the denominator and is
suppressed in the numerator as well as $-q^2$.
We use for $\alpha_s$ the following formula:

\begin{eqnarray}
\alpha_s(s) &=& \frac{1}{b \ln(-s/\Lambda_{QCD}^2)} =
\frac{1}{b \left[ \ln(s/\Lambda_{QCD}^2) - \imath\pi \right]}
\nonumber\\ &=&
\frac{1}{b} \left[ \frac{\ln(s/\Lambda_{QCD}^2)}
{\ln^2(s/\Lambda_{QCD}^2) + \pi^2} +
\frac{\imath\pi}{\ln^2(s/\Lambda_{QCD}^2) + \pi^2} \right]
\label{alpha},
\end{eqnarray}
where $b=(33-2n_f)/12\pi$ ($n_f$ -- the number of flavours).

Equation ~(\ref{alpha}) is written in such a way that $\alpha_s(s)$ has
non-zero
imaginary part when $s$ is positive.
Applying the transform Eq.~(\ref{invmellin})
to Eq.~(\ref{defB}) we
define

\begin{equation}
\label{r}
R^{(\pm)} = \frac{2}{\pi\omega}\int_0^{\infty} d\rho \exp(-\omega\rho) \Im_s
B^{(\pm)}(\rho) = \frac{A}{\omega},
\end{equation}
where $\rho=\ln(s/\mu^2)$,

\begin{equation}
A(\omega) = \frac{4C_F \pi}{b}
\left[ \frac{\eta}{\eta^2 + \pi^2} -  \int_0^{\infty}
\frac{d\rho \exp(-\rho\omega)}{(\rho + \eta)^2 + \pi^2}  \right],
\label{A}
\end{equation}
and we have introduced $\eta = \ln(\mu^2/\Lambda_{QCD}^2)$ assuming
$\mu >\Lambda_{QCD}$.

The first term in the square brackets in Eq.~(\ref{A}) corresponds to
the imaginary part of Eq.~(\ref{alpha}) and the second one comes from
 the imaginary part of $1/(s-\mu^2+\imath\epsilon)$.

The expression corresponding to graph~(b) in Fig. \ref{f0} is
\begin{equation}
\label{ladder}
 \frac{1}{8\pi^2}\left[ \frac{1}{\omega} + \lambda \right]
(F^{\pm}_0)^2 .
\end{equation}

Here, the softest partons
(i.e. partons with minimal $k_{\perp}$)
here are given by the intermediate quark pair. The remaining
graphs, ~(c) - (f)
in rhs of
Fig. \ref{f0}, correspond to the case when the softest parton
 is a gluon and therefore can be factorized out, i.e. its
propagator can be attached to the  external lines in all possible ways.

The amplitude $M_0$ in the lhs of Fig.~\ref{f0} is colourless,
i.e. it belongs to the
singlet representation of the group $SU_c(3)$.
As one of the virtual gluons is removed from the blobs in
graphs (c) - (f), these blobs are nolonger colourless.
They correspond to the colour octet scattering amplitude $M_8$,
where $k_{\perp}$ acts as a new IR cut-off. As   $M_8$ does not
depend on the longitudinal components of $k$, one can
integrate
over them, easily arriving at
\begin{equation}
\label{imcf}
\Im M^{\pm}_{(c-f)} = C_F
\int_{\mu^2}^{s} \frac{d(-k^2)}{(-k^2)}
\alpha_s(k^2)
\Re_s M_8^{\mp}(\frac{s}{-k_{\perp}^2}).
\end{equation}
Equation ~(\ref{imcf}) reads that there is total compensation
between the cuts of $M_8$, so that only cuts that do not
involve $M_8$ (i.e. two-quark ones) contribute
to $\Im M^{\pm}_{(c-f)}$.

As we are going to obtain
$M_0^{(-)}$ to all orders in $\alpha_s$ with SL accuracy, we must first
know $M_8^{(+)}$ with the same accuracy.  However, for an evaluation of
the asymptotic behaviour of $M_0^{(-)}$, this knowledge is not necessary.
Indeed, it was shown in ref. \cite{et} that in DLA the blobs in
Fig.~\ref{f0}c - f can be approximated, with a few percent accuracy, by
their Born contribution.  The reason is that the octet
amplitude $M_8^{(+)}$
decreases with energy so quickly that higher-loop
contributions can be neglected.  Motivated by this result and noticing
from the results of ref. \cite{egt} that, apart from running $\alpha_s$
effects, SL contributions do not change DL results drastically, we use
this approximation in the present work, replacing the blobs in
graphs (c) - (f) of the equation in Fig.~\ref{f0}  by
their Born value, although with the running $\alpha_s$.
Substituting the Born approximation $M_8^{Born}$ for the colour octet
amplitude $M_8$,

\begin{eqnarray}
\label{8born}
\Re M_8^{\mp}&\approx& \Re M_{8~Born}^{\mp} =
-2\pi \Re \left[
\alpha_s(s)\frac{s}{s - \mu^2 + \imath\epsilon} \mp
\alpha_s(-s)\frac{-s}{-s - \mu^2 + \imath\epsilon} \right]
\nonumber \\ &\approx& -2\pi \left[\Re \alpha_s(s) \mp
\Re \alpha_s(-s) \right],
\end{eqnarray}
we then obtain

\begin{equation}
\label{imcfborn}
\Im M_{(c-f)}^{(\pm)} =  -\frac{2\pi C_F}{2N}
\left[\Re \alpha_s(s) - \Re \alpha_s(-s)  \right]
\int_{\mu^2}^{s} \frac{d(-k^2)}{(-k^2)}
\alpha_s(k^2) .
\end{equation}
Finally, applying the Mellin transform Eq.~(\ref{invmellin}) to
Eq.~(\ref{imcfborn}), and using Eq.~(\ref{alpha}) for $\alpha_s$, we
obtain,
for contribution of graphs~(c) - (f) in Fig. \ref{f0}:

\begin{eqnarray}\label{dpm}
D^{(\pm)}(\omega) =
\frac{2C_F}{\omega b^2 N} \int_0^{\infty} d\rho e^{-\omega\rho}
\ln\left( \frac{\rho + \eta}{\eta}\right)
\left[ \frac{\rho + \eta}{(\rho + \eta)^2 + \pi^2} \mp
\frac{1}{\rho + \eta}
\right].
\end{eqnarray}

In DLA, when $\alpha_s$ is fixed, instead of Eq.~(\ref{dpm}) one gets

\begin{equation}
\label{ddla}
D^{(+)}_{DL}(\omega) = 0~~,
\qquad D^{(-)}_{DL}(\omega) =-\frac{4\alpha_s^2 C_F}{\omega^3 N}~~.
\end{equation}

The positive-signature contribution $D^{(+)}$ in the DLA is zero, so that
$M_8$ does not contribute to $M_0^{(+)}$ at all.
In the Feynman gauge this means that the
positive-signature contribution of non-ladder graphs in the DLA is zero
and only ladder graphs must be considered.  This remarkable observation
was first made in ref. \cite{gln} for QED processes, and later in
\cite{kl} for quark scattering. Equation ~(\ref{dpm})
shows that running $\alpha_s$ effects violate the total
compensation of non-ladder graph contributions. In our previous
work \cite{egt} we had neglected this non-compensation. We discuss it in
more detail in Section 4.

Now we are able to  write the full equation for
$F_0^{(\pm)}$ shown in Fig.~\ref{f0}, which
generalizes the equation for $F_0=F_0^{(+)}$ obtained in our earlier work
\cite{egt}:

\begin{equation}\label{equark}
F^{(\pm)}_0(\omega) = \frac{A(\omega)}{\omega} + \frac{1}{8\pi^2}
\left[\frac{1}{\omega} + \lambda\right]
\left(F^{(\pm)}_0(\omega)\right)^2 + D^{(\pm)}(\omega),
\end{equation}
where $D^{(+)}$ and $D^{(-)}$ are given by Eq.~(\ref{dpm});
$\lambda = 1/2$ corresponds to SL contributions not related to running
coupling effects.

Equation ~(\ref{equark}) has the following solution

\begin{equation}
F^{(\pm)}_0 =
4\pi^2~\frac{\omega-\sqrt{\omega^2 -
(1 + \lambda\omega) (A(\omega) + \omega D^{(\pm)}(\omega)) / 2\pi^2}}
{1 + \lambda\omega}~~.
\label{F0}
\end{equation}

Combining this result with Eqs.~(\ref{ims}), (\ref{defns}), (\ref{cir}) we
finally obtain the following expressions for the structure functions
$f_1^{NS}$ and $g_1^{NS}$:

\begin{eqnarray}\label{evir}
f_1^{NS} &=& \int_{-\imath \infty}^{\imath \infty}
\frac{d \omega}{2\pi \imath}C\left( \frac{1}{x}\right)^{\omega}
\omega \exp\left( \left[ (1+\lambda\omega) F_0^{(+)}/8\pi^2 \right]
y \right)
\nonumber \\
g_1^{NS} &=& \int_{-\imath \infty}^{\imath \infty}
\frac{d \omega}{2\pi \imath}C\left( \frac{1}{x}\right)^{\omega}
\omega \exp\left( \left[ (1+\lambda\omega) F_0^{(-)}/8\pi^2 \right]
y \right)
\end{eqnarray}
with $C$ arbitrary.

\section{Asymptotics of the non-singlet structure functions}

The small-$x$ asymptotic behaviour of $g_1^{NS}$ and $f_1^{NS}$ can be
obtained by evaluating the expressions found at the end of last section
with the saddle-point method. However, it is much easier to
get it directly, by noticing that at small $x$
$g_1^{NS}\sim x^{-\omega^{(-)}_0}$ and
$f_1^{(NS)} \sim x^{-\omega^{(+)}_0}$,
where $\omega_0^{(\pm)}$ is the rightmost singularity of $F^{(\pm)}$.
Equation ~(\ref{F0}) reads that this singularity is a square root branch
point given by a solution to\footnote{There is a misprint  in the
coefficient before the square brackets of Eq.~(52) in ref. \cite{egt}.
The true coefficient is $(2C_F/\pi b)$. However the numerical results of
ref.
\cite{egt} are correct.}

\begin{eqnarray}\label{master}
&&\omega^2 - (1 + \lambda \omega) \left\{ \left( \frac{2C_F}{\pi b}
\right) \left[ \frac{\eta}{\eta^2 + \pi^2} - \int_0^{\infty}
\frac{d\rho e^{-\omega\rho}}{(\rho + \eta)^2 + \pi^2} \right] +
\right.\nonumber \\ && \left.
\left(\frac{2C_F}{\pi b}\right)^2\frac{1}{4NC_F}
\int_0^{\infty} d\rho e^{-\omega\rho}
\ln\left( \frac{\rho + \eta}{\eta}\right)
\left[ \frac{\rho + \eta}{(\rho + \eta)^2 + \pi^2} \mp
\frac{1}{\rho + \eta}
\right]\right\}
 = 0~~.
 \end{eqnarray}
We recall  that $\rho = \ln(s/\mu^2)$ and $\eta =
\ln(\mu^2/\Lambda^2_{QCD})$.  This equation can be
solved numerically. Before doing so, let us notice that if we keep
$\alpha_s$ fixed in Eq.~(\ref{master}) and, for self-consistency, put
$\lambda = 0$, it transforms into the well-known algebraic equation in
the DLA:

\begin{equation} \label{eqasomega}
\omega^2 - \frac{2\alpha_sC_F}{\pi} -
\left(\frac{2\alpha_sC_F}{\pi}\right)^2 \frac{1}{4NC_F}
\frac{1}{\omega^2}[1 - (\pm 1)]  = 0,
\end{equation}
with the obvious solutions

\begin{equation} \label{omegadl}
\widetilde{\omega}^{DL}_{(-)} = \widetilde{\omega}^{DL}_{(+)}
\sqrt{ \frac{1}{2}\left[ 1 +
\left( 1 + \frac{4}{(N^2 - 1)} \right)^{1/2} \right]}\approx
1.055~\widetilde{\omega}^{DL}_{(+)},
\end{equation}
where the positive-signature leading singularity reads:

\begin{equation}
\label{omegaplusdl}
\widetilde{\omega}^{DL}_{(+)} = \sqrt{2\alpha_s^{DL} C_F/\pi}~~.
\end{equation}
Now let us come back to Eq.~(\ref{evir}).
Substituting the value of $F_0^{(\pm)}$ of Eq.~(\ref{F0}) at the point
$\omega = \omega_0^{(\pm)}$, where the square root turns to
zero, we arrive at
the power-like asymptotics:

\begin{eqnarray}
f_1^{NS} &\sim&
\left(\frac{1}{x}\right)^{\omega_0^{(+)}}
\left(\frac{Q^2}{\mu^2}\right)^{\omega_0^{(+)}/2},
\nonumber \\
g_1^{NS} &\sim&
\left(\frac{1}{x}\right)^{\omega_0^{(-)}}
\left(\frac{Q^2}{\mu^2}\right)^{\omega_0^{(-)}/2}~~.
\label{fasy}
\end{eqnarray}

\begin{figure}
\begin{center}
\begin{picture}(320,220)
\put(0,20){
\epsfbox{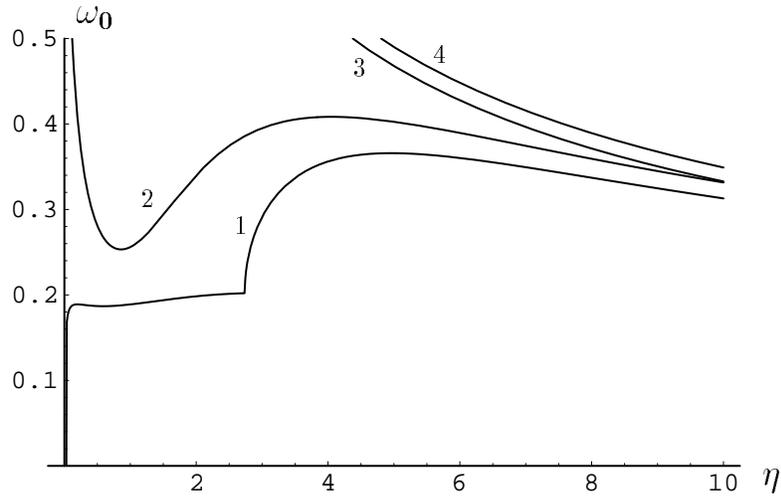}
}
\end{picture}
\end{center}
\caption{
Dependence of the intercept $\omega_0$ on infrared cut-off
$\eta=\ln(\mu^2/\Lambda_{QCD})$. 1: for $f_1^{NS}$, 2: for
$g_1^{NS}$, 3 and 4: for $f_1^{NS}$ and $g_1^{NS}$, respectively,
without accounting for $\pi^2$-terms.
}
\label{intercept}
\end{figure}

Obviously, the solutions $\omega_0^{(+)}$ and $\omega_0^{(-)}$ to
 Eq.~(\ref{master})
(also called the intercepts  of $f_1^{NS}$ and $g_1^{NS}$, respectively)
depend  on the choice of the parameters $n_f$, $\mu$,
$\Lambda_{QCD}$.  We recall that we keep $\mu >
\Lambda_{QCD}$. Also, $\mu$ must be greater than
the mass of the heaviest involved quark, which fixes $n_f$.
In numerical estimates below, we have used $\Lambda_{QCD}=0.1\GeV$
and $n_f = 3$. Equation ~(\ref{master}) has been solved numerically for
different values of $\mu$.
The dependence of the intercepts $\omega_0^{(\pm)}$ on $\mu$ (in terms of
$\eta$) is plotted  in Fig.~\ref{intercept}. The plot shows that
$\omega_0^{(-)}$ is greater than $\omega_0^{(+)}$ for all $\mu$.
Both $\omega_0^{(+)}$  and $\omega_0^{(-)}$
first grow with increasing  $\mu$, getting maximal at $\mu\approx
1\GeV$ , where

\begin{equation} \label{Omegaplus}
\omega_0^{(+)}\equiv\Omega^{(+)} = 0.37
\end{equation}
 and

\begin{equation}
\label{Omegaminus}
\omega_0^{(-)}\equiv\Omega^{(-)} = 0.4,
\end{equation}
and smoothly fall off after that.
Equation ~(\ref{master}) also tells us that contrary to the
double-logarithmic
approximation of Eq.~(\ref{omegadl}), there is no total compensation
between the DL contributions of the non-ladder
graphs to $\omega_0^{(+)}$ (see expression in the second squared
brackets)
when $\alpha_s$ is running. That fact was neglected in \cite{egt}.
Once taken into account, it leads to a steep increase of
both $\omega_0^{(+)}$  and $\omega_0^{(-)}$ at small $\eta$.  However,
we regard this increase as an artefact of our approach:
using expression
(\ref{alpha}) for  $\alpha$ is becoming unreliable at such
small $\eta$.
Another reason for not considering the small-$\eta$ region
is that when $\eta\leq 2.86$ the values of
$\omega_0^{(+)}$  become complex, which leads to
negative $f_1^{NS}$.  Thus, we find that our
approach is valid for  $\eta\geq 2.8$. Recalling that
$\eta =\ln[(\mu/\Lambda_{QCD})^2]$, with $\mu$ acting as an input
for our evolution equations, we see that,
for self-consistency, we must keep
 $\mu\ge\mu_{min} = 17.5\Lambda_{QCD} = 1.75\Gev$ as
 $\Lambda_{QCD}$ = 0.1\Gev. A further look at
 Fig.~\ref{intercept} shows that
  $\omega_0^{(+)}$  and $\omega_0^{(-)}$   are maximal at
$\eta\approx 5$. Then they smoothly  $(\sim 1/\eta)$ fall with
 growing $\eta$.  The reason for this fall-off is quite clear: at
large $\eta$,  the running $\alpha_s$ is approximated by
$1/(b \eta)$.  The $\eta$-dependence of
$\omega_0^{(\pm)}$, plotted in Fig.~\ref{intercept},
has such a complicated form because of the competition between the
relative weights of the $\pi^2$- and $\eta$-contributions
in the denominators of Eq.~(\ref{master}). In the absence of the
$\pi^2$-terms
$\omega_0^{(\pm)}$ will depend on $\eta$ in a smoother way, as shown
by the curves (3) and (4) in Fig.~\ref{intercept}. We will further discuss
this plot in the next section.

 \section{Comparing with the leading logarithmic approximation}
The power-like (Regge-like) small-$x$ behaviour of
$f_1^{NS}$ and $g_1^{NS}$, Eq.~(\ref{fasy}),
with DL intercepts $\omega^{(\pm)}_0=\widetilde{\omega}^{DL}_{(\pm)}$
given by Eqs.~(\ref{omegadl}) and (\ref{omegaplusdl}),  was obtained in
refs. \cite{emr} and \cite{ber}.  These expressions explicitly include
the dependence of $\alpha_s$ upon an unknown scale, because the DL
approximation treats $\alpha_s$
as fixed. On the other hand, the LLA approach postulates that
asymptotically, at very high energies, LLA results should dominate
over all sub-leading contributions. Having accounted for the
SL contributions, we are now able to check the validity
of the LLA predictions for the NS structure
functions. First, let us notice that accounting for running
$\alpha_s$ does not lead to intercepts $\omega_0^{(\pm)}$
explicitly dependent on $\alpha_s$, although they depend on
$\Lambda_{QCD}$. Then, thanks to our choice of the
infrared regulation, Eq.~(\ref{mu}), $\omega_0^{(\pm)}$  depend on
$\mu$. But in spite of such a difference between
$\omega_0^{(\pm)}$  and the DL intercepts
$\widetilde{\omega}^{DL}_{(\pm)}$, one can see from
Eq.~(\ref{omegadl}) and Fig.~\ref{intercept} that the DL ratio
$\left[\widetilde{\omega}^{DL}_{(-)} -
\widetilde{\omega}^{DL}_{(+)}\right]/
\widetilde{\omega}^{DL}_{(+)} = 0.055$
basically holds for $\omega_0^{(\pm)}$ also, changing
slightly from 0.09 at $\eta =3$ to 0.06 at $\eta=8$.
Therefore, the DL result for the ratio in Eq.~(\ref{omegadl})
is a good   approximation. The
$\pi^2$-terms that we have accounted for
in Eq.~(\ref{master})are quite important because of
their large value. In principle, they are
formally beyond the SL accuracy we have kept through the paper.
Indeed, contributions $\sim\pi^2$ may appear also from integrations
over phase space.  Such contributions in higher orders of the perturbative
series are beyond the control of any kind of logarithmic approximation.
On the other hand the $\pi^2$ we have accounted for appear as the result
of respecting
the analytical properties of the scattering amplitudes. So, we cannot
simply neglect them.
Moreover, it has been shown recently in ref. \cite{kataev} that a
value $\Omega^{(-)} = 0.4$ is in good agreement with the
experimental data for the structure function $F_3$ . When the $\pi^2$ in
Eq.~(\ref{master}) are dropped,  $\Omega^{(-)} = 0.4$ corresponds to
$\mu \approx 5.5\GeV$ (see curves 3 and 4 in Fig.~\ref{intercept}).
At this value of $\mu$ the SL contributions not related to the running
$\alpha_s$ are small, i.e. one can neglect them, putting $\lambda = 0$
in Eq.~(\ref{master}). Therefore we conclude that $\mu_0 \equiv
5.5\GeV$ is a good estimate of the value of the mass scale for LLA
evolution equations. In other words, $\mu_0^2$ acts as
a momentum scale for the evolution
equations obtained in LLA. Therefore these equations are not
expected to reproduce the correct $Q^2$-dependence until at least

\begin{equation}\label{qmu}
Q^2>\mu_0^2 \approx 30 \Gev^2~~.
\end{equation}
As $\mu_0^2$ is pretty close to the typical values of $Q^2$ at HERA,
we can see that using the LLA evolution equations  is not very reliable
for such $Q^2$.

Let us discuss the $x$-dependence. In refs. \cite{b,jk,kk}, it is assumed
that $\alpha_s$ depends on $Q^2$ in the expression for the intercepts.
On the other hand the results of ref. \cite{egt} show that there is
no such dependence. Now we have seen that in LLA $\alpha_s$
should depend rather
on $\mu^2_0$ than on $Q^2$, and as the $Q^2$ range covered at HERA is not
far from
$\mu^2_0$ the estimate $\alpha_s=\alpha_s(Q^2)$ might still be in
reasonable
agreement with the data.

\section{Comparison with  DGLAP}

Now let us compare our results
with the DGLAP results for the non-singlet structure
functions. First, let us check that
the corresponding DGLAP evolution equations can be obtained
from Eq.~(\ref{eqir}). Writing Eq.~(\ref{eqir}) as

\begin{equation}\label{eqir2}
\frac{\partial}{\partial y} F^{(\pm)}(\omega, y)  =
\left[-\omega + \frac{1}{8\pi^2}(1 + \lambda \omega)\right]
F^{(\pm)}_0(\omega)
F^{(\pm)}(\omega, y)
\end{equation}
and noting that the term $-\omega$ in the square
brackets is absorbed by changing
the Mellin factor $(s/\mu^2)^{\omega}$  to
$x^{-\omega}$ (see Eq.~(\ref{evir})), then

\begin{equation}
\label{g}
\widetilde{\gamma}^{(\pm)} =
\frac{1}{8\pi^2}(1 + \lambda \omega)
 F^{(\pm)}_0(\omega)
\end{equation}
are the new small-$x$ non-singlet anomalous dimensions.
However, when $x \sim 1$
the main contribution of the Mellin factor $\exp [\omega \ln (1/x)]$
comes from the region of rather large values of $\omega$, where
$\omega \ln(1/x) \leq 1$.  That makes it possible to expand the exponent
$F^{(\pm)}_0$ in Eq.~(\ref{evir}) into a series in
$1/\omega$. Doing so,
we obtain

\begin{equation}\label{gammaser}
\widetilde{\gamma}^{(\pm)} \approx (1+\lambda\omega)
\frac{F^{(\pm)}_0}{8\pi^2} =
\frac{A}{8\pi^2} \left[\frac{1}{\omega} + \lambda \right] +
\left(\frac{A}{8\pi^2}\right)^2 \frac{1}{\omega} \left[
\frac{1}{\omega} + \lambda \right]^2  + D^{(\pm)} + O(A^3)~~.
\end{equation}
Retaining only the first  term
in the rhs of Eq.~(\ref{gammaser}), one arrives at
the expression

\begin{equation}
\label{gamma1}
\widetilde{\gamma}^{(\pm)}_{(1)} =  \frac{A}{8\pi^2}
\left[\frac{1}{\omega} + \lambda \right]
\end{equation}
for the leading order anomalous dimension, which is similar to
the expression for the leading order
DGLAP non-singlet anomalous dimension

\begin{equation}
\label{gamma1ap}
\gamma_{(1)}  =
\frac{\alpha_s(Q^2)C_F}{2\pi}
\left[\frac{1}{\omega} + \lambda \right]~~.
\end{equation}

Let us first compare the two results by  assuming that $\alpha_s$ is
fixed. Equations ~(\ref{gamma1}) and (\ref{gamma1ap})  coincide
completely because in this case $A= 4\pi\alpha_sC_F$. Let us then
compare the next two terms in Eq.~(\ref{gammaser}) with the second
order DGLAP anomalous dimension $\gamma^{(\pm)}_{(2)}$, again for fixed
$\alpha_s$.  Retaining the leading terms, proportional to
$1/\omega^3$ , we can see that

\begin{equation}
\label{gamma2}
\widetilde{\gamma}^{(\pm)}_{(2)} = \left(\frac{A}{8\pi^2}\right)^2
\frac{1}{\omega} \left[\frac{1}{\omega}+\lambda\right]^2 + D^{(\pm)}
\end{equation}
transform into (see Eq.~(\ref{ddla}))

\begin{equation}
\label{gamma2dlplus}
\gamma^{(+)}_{(2)} = \left(\frac{\alpha_s}{\pi} \right)^2
 \frac{C^2_F}{2}  \frac{1}{\omega^3}
\end{equation}
and
\begin{equation}
\label{gamma2dlminus}
\gamma_{(2)}^{(-)} = \left(\frac{\alpha_s}{\pi} \right)^2
\left[ \frac{C^2_F}{2} + \frac{C_F}{N} \right] \frac{1}{\omega^3},
\end{equation}
which coincide with the leading terms
of Eq.~(B18) of ref.\cite{fkl}, also used in ref. \cite{grsv} for the
second
order non-singlet anomalous dimensions. Unfortunately,
we do not obtain exact coincidence between non-leading terms, which are
proportional to
$1/\omega^3$, but this can be explained by a different choice of the
factorization and regularization procedures. Therefore, the
non-singlet anomalous dimensions  given by Eq.~(\ref{g})
correspond to a
resummation of $\gamma_{(1)}$ to all orders in the QCD coupling and
to accounting for the signature-dependent DL
contributions to order $\alpha_s^2$. Such a resummation
leads to the power-like small-$x$ behaviour, which cannot be achieved
by incorporating any finite number of NLO contributions into
expressions for DGLAP anomalous dimensions.

Another  difference between IREE and DGLAP is the different
treatment of the QCD coupling. Although both approaches use the same
formula (Eq.~(\ref{alpha})) for $\alpha_s$,  the DGLAP uses
$\alpha_s = \alpha_s(k^2_{i \perp})$ in $i$-th ladder rung,
$k_{i \perp}$ being the ladder quark transverse
momentum whereas IREE suggests that in the $i$-th rung
$\alpha$ should depend on the gluon virtuality~\footnote{We use here
the standard numeration for the momenta of the virtual ladder quarks
$k_i$,
so that $i = 0$ corresponds to $k_0 = p$ and, increasing the number of
runs, it leads to the
top of the ladder.} $(k_i -k_{i-1})^2$ .
Let us explain this difference.
When $x \neq 1$, the Born contribution to $f^{NS}$ is zero and
$f^{NS}$ obeys
the Beth - -Salpeter equation

\begin{equation}
\label{dglap}
f^{NS} =
\int d^4 k \Phi((q + k)^2) \frac{1}{k^2} \Im
\frac{\alpha_s((p-k)^2)}{[(p - k)^2 ]}~~.
\end{equation}
We have suppressed all unimportant factors in Eq.~(\ref{dglap})
and have used the
notation $k$ instead of $k_1$ for the ladder quark momentum in the
lowest rung. We have noted as $\Phi$ for off-shell $f^{NS}$.
Usually it is convenient to integrate over $k$ in
(\ref{dglap}), using the Sudakov parametrization:
\begin{equation}
\label{sudak}
k_{\nu} = \alpha (q + xP)_{\nu} + \beta p_{\nu} + k_{\perp},
\end{equation}
so that
\begin{equation}
\label{ab}
2(pk) = 2(pq)\alpha , \quad 2(qk) = 2(pq) (\beta - x\alpha) , \quad
k^2 = 2(pq)\alpha\beta  - k^2_{\perp}~~.
\end{equation}
But here it is more convenient to use the variable

\begin{equation}
\label{m2}
m^2 \equiv (p - k)^2
\end{equation}
instead of the Sudakov variable $\alpha$, writing Eq.~(\ref{dglap}) as

\begin{eqnarray}
\label{bm2}
f^{NS} = \int d^2k_{\perp}d\beta dm^2 &\Phi&
\left(\left[ s(\beta - x)(1 - \beta) - m^2(\beta - x)
- k^2_{\perp}(1 - x)\right]
(1 - \beta)^{-1}\right)~\cdot \nonumber\\   &\Im&
\frac{\alpha_s(m^2)} {[\beta m^2 + k^2_{\perp}][m^2]}~~.
\end{eqnarray}
The argument of $\Phi$ in Eqs.~(\ref{dglap}) and (\ref{bm2})
should be positive, i.e.

\begin{equation}
\label{argument}
\left[ s(\beta - x)(1 - \beta) - m^2(\beta - x)
- k^2_{\perp}(1 - x) \right] \ge 0~.
\end{equation}
When $x \sim 1$, one can neglect the last term in Eq.~(\ref{argument}).
It leads to $(\beta - x)(s(1 - \beta) - m^2) \ge 0$, which
allows $m^2$ to be neglected. Therefore, instead of Eq.~(\ref{bm2}),
we obtain

\begin{equation}
\label{bmk}
f^{NS} =
\int d^2k_{\perp}d\beta dm^2 \Phi\left(
\left[ s(\beta - x)(1 - \beta) \right]
(1 - \beta)^{-1}\right)
\Im\frac{\alpha_s(m^2)}
{[\beta m^2 + k^2_{\perp}][m^2]}~~.
\end{equation}
After that, one can apply the Cauchy theorem relating an analytic
function to its imaginary part. This permits to perform an integration
over
$m^2$ by taking residues. As $m^2$ is positive, the only residue is
$m^2 = - k^2/\beta$. Thus, we  obtain

\begin{equation}
\label{bk}
f^{NS} =
\int d^2k_{\perp}d\beta  \Phi\left(
\left[ s(\beta - x)(1 - \beta) \right]
(1 - \beta)^{-1}\right)
\frac{\alpha_s(-k^2_{\perp}/\beta)}
{ k^2_{\perp}}~~.
\end{equation}
As $\beta \ge x$ and $x \sim 1$, one can drop $\beta$ in the
argument of $\alpha_s$ and arrive at

\begin{equation}
\label{bk2}
f^{NS} =
\int d^2k_{\perp}d\beta  \Phi\left(
\left[ s(\beta - x)(1 - \beta) \right]
(1 - \beta)^{-1}\right)
\frac{\alpha_s(-k^2_{\perp})}
{ k^2_{\perp}}~~,
\end{equation}
which corresponds to the DGLAP.
Usually the negative sign of the argument of $\alpha_s$ is dropped
together with the negative sign of the argument in Eq.~(\ref{alpha}).
As the argument of $\alpha_s$ in Eq.~(\ref{bk}) is negative, the
expression for $\alpha_s$ in the DGLAP approximation of Eq.~(\ref{bk2})
does not contain $\pi^2$-terms at all, while the argument of
$\alpha_s$ in Eq.~(\ref{dglap}) is positive and therefore there are
$\pi^2$-terms in
Eq.~(\ref{dglap}).

Obviously, the transition from Eq.~(\ref{dglap})
to Eq.~(\ref{bk}) holds only when $x \sim 1$, which is the DGLAP
kinematics, and fails for small $x$. In particular, neglecting
$k^2_{\perp}$ in Eq.~(\ref{argument}) has an important consequence for
the DGLAP evolution, where one  picks up logarithmic
contributions through integrations over $k_{\perp}$.  At small $x$ one
should use the $\alpha_s$-dependence given by Eq.~(\ref{A}) rather
than that given by Eq.~(\ref{bk2}).

\section{Conclusions}

In conclusion, we
have obtained a generalization of the second-order
DGLAP evolution equations for the non-singlet structure functions
at small $x$. Besides resumming the LO DGLAP  anomalous
dimension to all orders in
$\alpha_s$ and  accounting for running $\alpha_s$ effects, we
account for the difference between $f_1^{NS}$ and $g_1^{NS}$, which is
due to
the difference in the second-order anomalous dimensions.
We have shown that, with
single-logarithmic contributions and  running $\alpha_s$ taken
into account, $f_1^{NS}$ and $g_1^{NS}$ have
the scaling-like asymptotic behaviour

\begin{equation}
 \sim x^{- \omega_0^{(\pm)}}
\left(\frac{Q^2}{\mu^2}\right)^{(\omega_0^{(\pm)}/2)},
\label{fin}
\end{equation}
as also obtained earlier in the DLA  \cite{emr,ber}
at asymptotically small $x$. We have
appropriately calculated the intercepts
$\omega_0^{(\pm)}$.
We have made detailed comparisons of our results with the predictions
obtained in the leading-logarithmic approximation and with
the DGLAP.
Results obtained with  the evolution equations
usually depend on a mass scale
$\mu^2_0$ acting as a starting point of the evolution
with respect to $Q^2$, and
we estimate the value for this mass scale for a  LLA evolution
equation as $\mu^2_0\approx 30\Gev^2$. This implies that these
equations can reproduce the correct $Q^2$-dependence for
the non-singlet structure functions (and likely, for the
singlet ones) only for approximately $Q^2 > 30\GeV^2$ .
On the other hand, as far as the $x$-dependence is concerned, we have also
shown that the estimate
$\alpha_s=\alpha_s(Q^2)$ used in refs. \cite{b,jk,kk} in
LLA can be in a reasonable agreement with the  $x$-behaviour of
the structure functions for the HERA data.

\section{Acknowledgements}
The work is supported in part by grant INTAS-97-30494, and by EU QCDNET
contract FMRX-CT98-0194.


\end{document}